\documentclass{appolb}
\usepackage{graphicx}

\begin{document}
\title{ Holographic picture of heavy vector mesons in a finite density plasma %
\thanks{Presented at Excited QCD 2017 }%
}
\author{Nelson R. F. Braga and  Luiz F.  Ferreira 
\address{ Instituto de F\'{\i}sica,
Universidade Federal do Rio de Janeiro, Caixa Postal 68528, RJ
21941-972 -- Brazil  }
 }
\maketitle
\begin{abstract}
We present the results of  a holographic model for heavy vector mesons   inside a plasma at finite temperature and density.  The spectral function shows a nice description of the dissociation of the mesons in the medium, corresponding to the decrease in the height of the peaks as the temperature and (or) the density increase.  We consider the case of bottomonium states at finite temperature and also the case of charmonium dissociation at zero temperature but finite chemical potential. 
 
\end{abstract}

  
\section{Introduction}
 
Understanding heavy meson dissociation inside a thermal medium can be a helpful tool for studying the properties of the quark gluon plasma formed in heavy ion collisions\cite{Matsui:1986dk} (see also \cite{Satz:2005hx}). 
A holographic description of the dissociation of charmonium and bottomonium states in a plasma at finite temperature but  zero density appeared recently in ref.\cite{Braga:2017oqw}. This work is an extension to the case of a finite density medium of the model presented in refs.\cite{Braga:2015jca,Braga:2016wkm}.  

The holographic model of ref.\cite{Braga:2015jca}  is an alternative version of the soft wall model that  is consistent with the observation that decay constants of hadronic resonances decrease with radial excitation level. This property is not reproduced in the hard wall  AdS/QCD model,  that appeared in refs. \cite{Polchinski:2001tt,BoschiFilho:2002ta,BoschiFilho:2002vd}, or in the original soft wall model\cite{Karch:2006pv}. 

One can understand the importance of having a consistent description of the decay constants by noticing  that,  at zero temperature, the two point function of hadronic currents has a spectral decomposition  in terms of masses and   decay constants  of the states:  
 \begin{equation}
\Pi (p^2)  = \sum_{n=1}^\infty \, \frac{f_n^ 2}{(- p^ 2) - m_n^ 2 + i \epsilon} \,. 
\label{2point}
\end{equation} 
 At finite temperature the particle content of a theory is described by the spectral function. 
The quasi-particle states appear as peaks that decrease as the temperature or the density of the medium increase.  The zero temperature and density  limit of the spectral function corresponds to the imaginary part of eq. \ref{2point}. It is a sum of Dirac delta functions with coefficients proportional to the square of the decay constants. So, it is important to consider a model  consistent with the behavior of zero temperature decay constants in order find a reliable picture of the finite temperature case.  
  
 \section{Highlights of the holographic model} 
   
Heavy vector mesons are described by a  vector field $V_m = (V_\mu,V_z)\,$ ($\mu = 0,1,2,3$), assumed to be  dual of  the gauge theory current $ J^\mu = \bar{q}\gamma^\mu q \,$. The action is 
\begin{equation}
I \,=\, \int d^4x dz \, \sqrt{-g} \,\, e^{- \Phi (z)  } \, \left\{  - \frac{1}{4 g_5^2} F_{mn} F^{mn}
\,  \right\} \,\,, 
\label{vectorfieldactionzeroptemp}
\end{equation}
\noindent where $F_{mn} = \partial_mV_n - \partial_n V_m$ and $\Phi = k^2z^2   $ is the soft wall background, with the parameter  $k$  playing the role of an IR, or mass, energy scale. 
The space is a charged black hole 
\begin{equation}
 ds^2 \,\,= \,\, \frac{R^2}{z^2}  \,  \Big(  -  f(z) dt^2 + \frac{dz^2}{f(z) }  + d\vec{x}\cdot d\vec{x}  \Big)   \,,
 \label{MinkoviskyMetric}
\end{equation}
with
\begin{equation}
f (z) = 1 - \frac{z^ 4}{z_h^4}-q^2z_{h}^{2}z^4+q^2z^6 \,.
\end{equation}
The parameter $q$ is the charge of the black hole and  $z_h$ is the position of the horizon, defined by the condition $f(z_h)=0$.  They are related by 
\begin{equation} 
\mu=  q z_h ^{2} \,.
\label{c}
\end{equation} 
The gauge theory  correlators are calculated at  a finite position $ z = z_0$ . The inverse of the position is an  UV  energy  parameter.  More details can be found in ref.\cite{Braga:2017oqw}.

\section{Results}

 \begin{figure}[h]
\label{g6}
\begin{center}
\includegraphics[width=12.5cm]{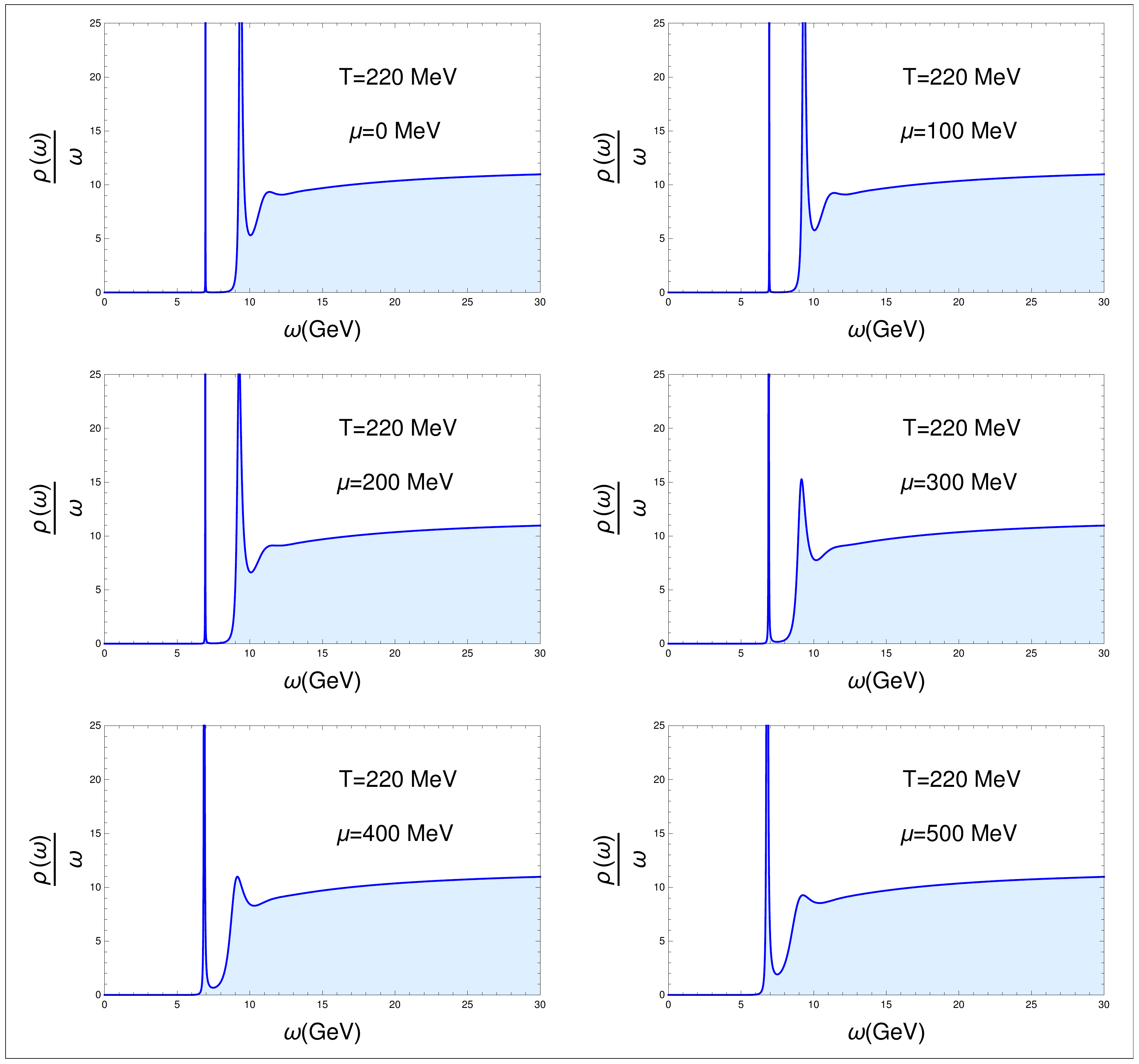}
 \end{center}
\caption{Spectral functions for bottomonium states with temperature $T=$ 220 MeV at 6 representative  values of $\mu$ }
\end{figure}
     
\begin{figure}[h]
\label{g6}
\begin{center}
\includegraphics[width=12.5cm]{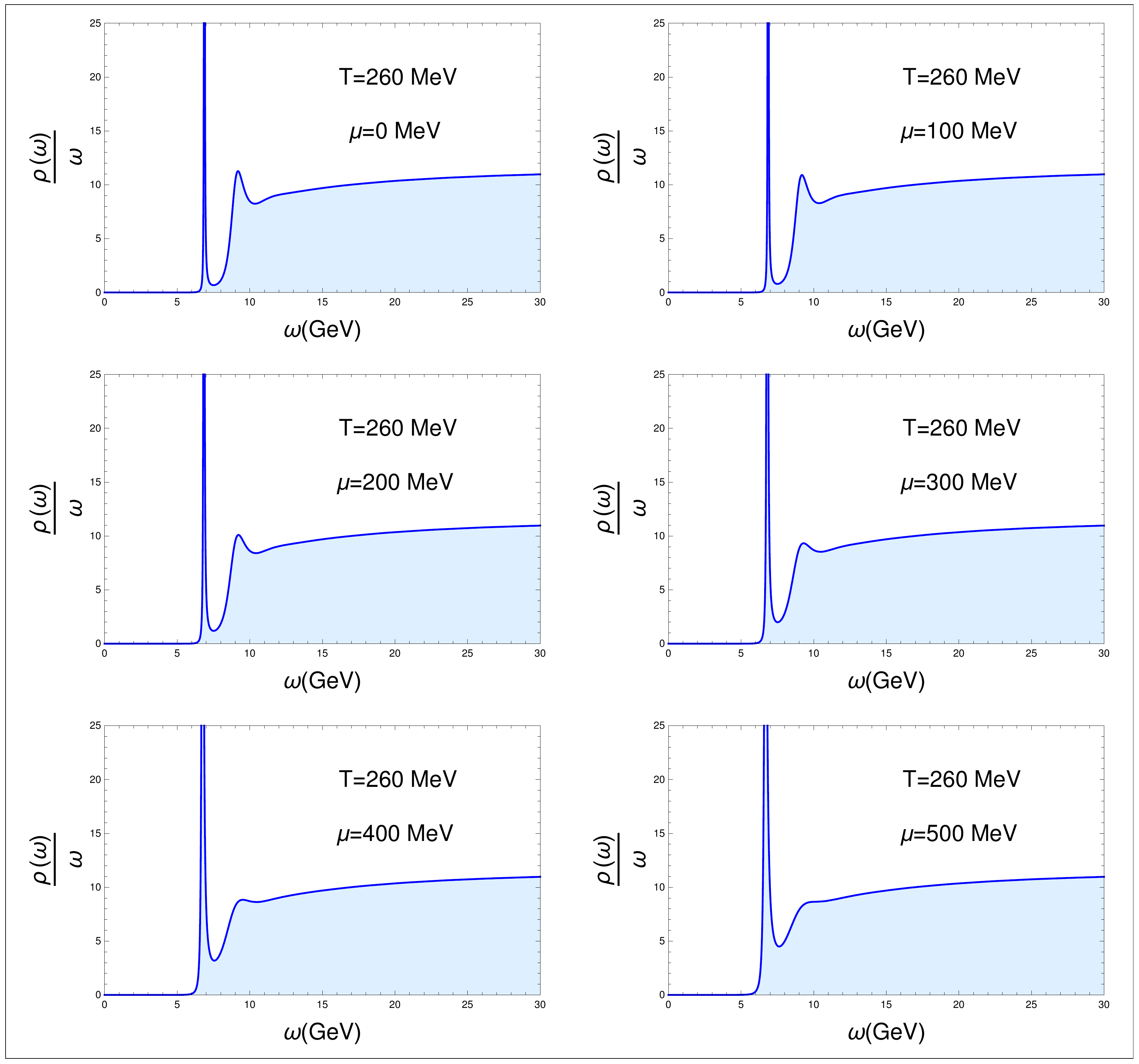}
\end{center}
\caption{Spectral functions for  bottomonium states with temperature $T=$ 260 MeV at 6 representative  values of $\mu$ }
\end{figure}

    The spectral function of  bottomonium  can be calculated using the membrane paradigm \cite{Iqbal:2008by}, for more details see ref. \cite{Braga:2017oqw}.  For the model parameters, we use the values  obtained in ref. \cite{Braga:2015jca} from the fit of the experimental data at zero temperature and chemical potential for bottomonium and charmonium:

\begin{equation}
 1/z_0 = 12.5  \,{\rm GeV} \,\,\, \,\,\,\,\,\,  ;  \,\,\,\,\,\,\  k_b = 3.4 \,  {\rm GeV  } \,\,  ;  \,\,\,\,\,\,\  k_c = 1.2 \,  {\rm GeV  } \,\,.
 \end{equation}

 The critical confinement/deconfinement temperature of this model -- with ultraviolet cut off -- was estimated for the case of zero chemical potential in  ref. \cite{Braga:2016wkm} as $T_c = 191$ MeV.
 
  In figure {\bf 1 } we show the spectral function for the temperature $ T $ = 220 MeV for six  different values of the chemical potential from $ \mu = 0 $ to $ \mu = 500 $ MeV.  One can see on the first  panel the peaks corresponding to the $1 S$ and $2S $ states and a very small peak corresponding to the $3S$ state. Then, raising the value of $\mu $ the second peak dissociates while the first one develops a larger width.  
  
Then in figure {\bf 2 } we show the case of $ T= 260 $ MeV where one sees a clear reduction of the second quasi-particle peak associated with the $2 S$ state at $\mu =0$. Increasing $\mu$ one sees that this state is completely dissociated in the medium at $\mu = 500 $ MeV.

\begin{figure}[h]
\label{g6}
\begin{center}
\includegraphics[width=12.5cm]{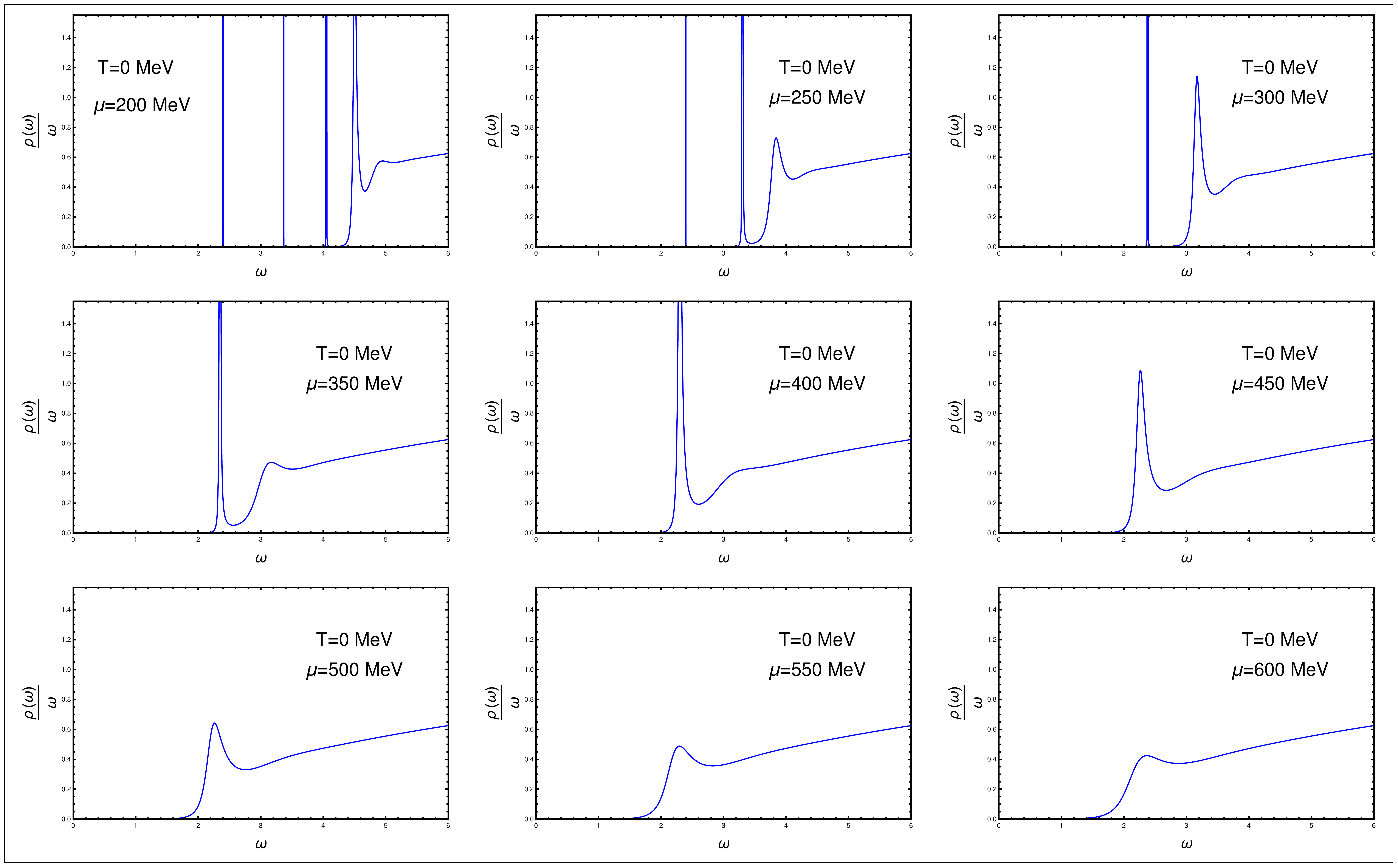}
\end{center}
\caption{Spectral functions for charmonium  states with vanishing temperature  at 8 representative  values of $\mu$ }
\end{figure}

One can also use this holographic model to look at the particular case of zero temperature and finite chemical potential. In figure  {\bf 3} we present the results for charmonium  from   $\mu = 200 $ to $\mu = 600 $ MeV  with temperature $T=0$ fixed.  One can see that at $\mu = 200$ there are  four peaks. Then increasing $\mu $ the quasi-particle peak decreasing, the dissociation of the 1S state happens at $\mu = 600 $ MeV.

 \section{Conclusions }

A nice picture for  the dissociation of bottomonium states emerged in ref. \cite{Braga:2017oqw} from the extension of the model of references \cite{Braga:2015jca,Braga:2016wkm} to finite chemical potential and temperature. The plots obtained  show the bottomonium dissociation as a function of density and temperature.  It was shown also that it is possible to analyse the spectral function at zero temperature and finite density.

\noindent {\bf Acknowledgments:}     N.B. is partially supported by ``Conselho Nacional de Desenvolvimento Cient\'{\i}fico e Tecnol\'ogico  (CNPq)'', Brazil, and L. F.  is supported by 
``Coordena\c{c}\~ao de Aperfeiçoamento de Pessoal de N\'{\i}vel Superior (Capes)'', Brazil.

\end{document}